\documentclass{article}
\usepackage{spconf,amsmath,graphicx}

\usepackage[utf8]{inputenc} 
\usepackage[T1]{fontenc}    
\usepackage{multirow}
\usepackage{amsfonts}       
\usepackage{nicefrac}       
\usepackage{graphicx}
\usepackage{caption}
\usepackage{subcaption}

\title{A Cycle-GAN Approach to Model Natural Perturbations in Speech for ASR Applications}
\name{Sri Harsha Dumpala, Imran Sheikh, Rupayan Chakraborty, Sunil Kumar Kopparapu}
\address{TCS Research and Innovation - Mumbai, INDIA}
\begin{document}
%
\maketitle
\begin{abstract}
Naturally introduced perturbations in audio signal, caused by emotional and physical states of the speaker, can significantly degrade the 
  performance of Automatic Speech Recognition (ASR) systems. In this paper, we propose a front-end based on Cycle-Consistent Generative Adversarial Network (CycleGAN) which transforms naturally perturbed speech into normal speech, and hence improves the robustness of an ASR system. The CycleGAN model is trained on non-parallel examples of perturbed and normal speech. Experiments on spontaneous laughter-speech and creaky-speech datasets show that the performance of four different ASR systems improve by using speech obtained from CycleGAN based front-end, as compared to directly using the original perturbed speech. Visualization of the features of the laughter perturbed speech and those generated by the proposed front-end further demonstrates the effectiveness of our approach.
\end{abstract}
\section{Introduction}
\label{Intro}

Performance of Automatic Speech Recognition (ASR) systems have seen significant jumps with the adoption of deep learning techniques.
Recently, ASR systems have been shown to perform on par with human transcribers \cite{ xiong2018microsoft}. At the same time, the use of voice assistants such as Siri, Google Assistant, Amazon Alexa etc., have led to the wide use of ASR systems in various day-to-day applications.
However, recent studies have shown that adversarial examples, generated by either adding a small amount of noise or by modifying a few bits of the audio signal, can be used to attack ASR systems to generate a completely different output \cite{iter2017generating, carlini2018audio},
even though the changes in the audio signal cannot be perceived by humans. 
Similar to these artificial perturbations in the audio signal, 
natural perturbations in human speech may also have an adverse effect
on the performance of ASR systems.
Natural perturbations in speech can arise due to the psychological and physical state of the speaker. Examples of naturally perturbed speech include expressive speech containing different emotions such as laughter, excitement, frustration, etc. and speech generated with different voice qualities such as creakiness, breath, etc.

In this paper, we show that the performance of the state-of-the-art deep neural network based ASR systems can significantly degrade for speech colored either by emotion or voice-quality. We show that these natural perturbations
can be handled by Cycle-consistent Generative Adversarial Networks (CycleGANs) \cite{zhu2017unpaired}, a variant of Generative Adversarial Networks (GANs) \cite{goodfellow2014generative}
which can learn distributions of data across different domains even without a parallel corpus.
The generator from our CycleGAN model learns to filter out the natural perturbations in speech and hence can be used as a front-end processor to improve the robustness of ASR to natural perturbations. Interestingly, in absence of these perturbations in the input speech to the CycleGAN model, the front-end processing does not affect the ASR performance.
The main contributions of this work are
\begin{itemize}
 \item An analysis of the performance of state-of-the-art ASR systems on naturally perturbed laughter and creaky-speech.
 \item An approach to train a CycleGAN model to obtain a front-end for transforming perturbed speech into normal speech.
 \item An analysis of the proposed front-end and its effectiveness in improving performance of state-of-the-art ASR systems.
\end{itemize}

The rest of the paper is organized as follows. Section \ref{RelatedWork} provides a brief overview of the related work. Detailed description of our CycleGAN model is given in Section \ref{CGANS}. Experiments and results are presented in Section \ref{exp} followed by an analysis on the learned transformation in Section \ref{analysis} and the conclusion in Section \ref{conc}. 

\begin{figure*}[!htb]
	\vspace{0.3cm}
	\centering
	\begin{subfigure}[b]{\textwidth}
		\centering
		\includegraphics[width=0.75\linewidth]{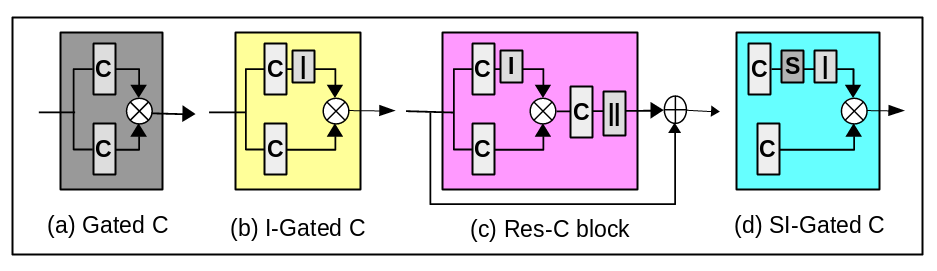}
		\caption{Different blocks used in generator and discriminator networks.}
		\label{fig:blocks}
	\end{subfigure}\hfill
	\begin{subfigure}[b]{\textwidth}
		\centering
		\includegraphics[width=0.8\linewidth]{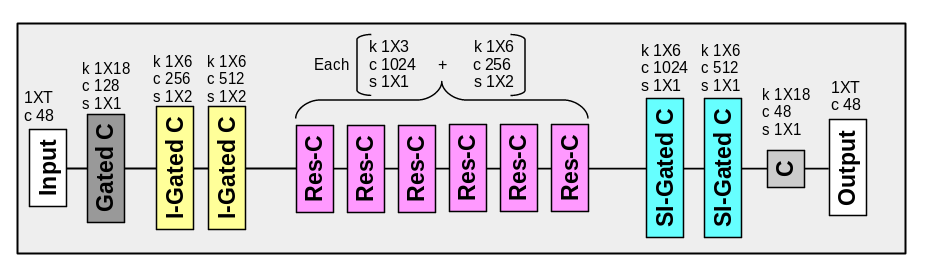}
		\caption{Block diagram of the generator network. Note: 'c' refers to channels, 'k' refers to kernel size and 's' refers to strides. 'T' refers to number of frames in the input.}
		\label{fig:Gen_block}
	\end{subfigure}\hfill
	\begin{subfigure}[b]{\textwidth}
		\centering
		\includegraphics[width=0.65\linewidth]{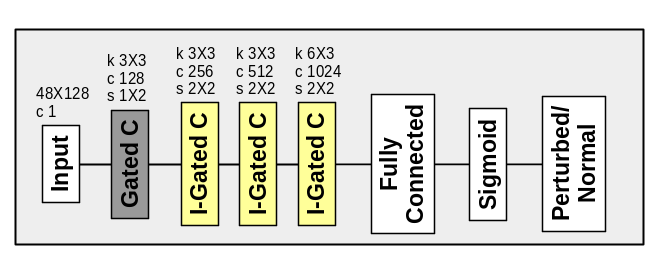}
		\caption{Block diagram of the discriminator network. Note: 'c' refers to channels, 'k' refers to kernel size and 's' refers to strides.}
		\label{fig:Disc_block}
	\end{subfigure}
	\caption{Block diagram of our proposed Cycle-GAN model to transform perturbed speech to normal speech. Note: in above figure, C refers to convolutional layer, Gated-C is gated convolution, I-Gated-C is instance normalized gated convolution, Res-C is residual convolution block and SI-Gated-C is pixel shuffled I-Gated-C}
	\label{fig:CGAN_Architecture}
\end{figure*}

\section{Related Work}
\label{RelatedWork}
Previous work have analyzed the effect of emotional speech on ASR and shown significant degradation in the performance of GMM-HMM-based ASR systems \cite{athanaselis2005asr, vlasenko2012towards}. They proposed adaptation of the acoustic and language models of the ASR system to capture the variations exhibited by emotive speech, in order to improve the ASR performance.
As opposed to model adaptation, we propose an approach based on transforming emotional speech to normal speech.
Recently, emotive-to-neutral speech conversion has been achieved by modeling prosody-based features \cite{raju2016application}. But this approach requires a parallel corpus (i.e., same utterance spoken in neutral and with emotion), which is very difficult to collect for spontaneous speech. 
Similarly, 
GMM-HMM-based systems have been considered for synthesizing creaky-speech \cite{narendra2017generation}, but no previous work has considered the conversion of creaky to neutral speech 
due to lack of a parallel corpus of creaky and neutral speech. 

We propose a parallel-data-free approach to 
transform speech perturbed with emotions and voice quality to normal speech, based on CycleGANs \cite{zhu2017unpaired}.
CycleGAN was earlier used for voice conversion without parallel-data \cite{kaneko2017parallel}.
Compared to \cite{kaneko2017parallel}, our approach provides a front-end processor which can add robustness to ASR on utterances perturbed with emotion and voice quality.
This paper presents the details of our CycleGAN model, the training loss functions and additional experimental results which further validate the performance of our approach.

\section{Perturbed Speech to Normal Speech Transformation with CycleGANs}
\label{CGANS}
GANs consist of two different networks i.e., a generator $G$ and a discriminator $D$. Generator is used to generate the fake samples $G(z)$, that resemble a given data distribution $X$, by taking random sample $z$ from a prior distribution $p_z$ as input, and the discriminator is used to discriminate fake samples from real samples in the data $X$. Both, generator and discriminator are trained using an adversarial loss function  \cite{goodfellow2014generative}.
GANs 
were initially proposed for the generation of images when provided with some arbitrary random noise as input, and thereafter 
have achieved impressive results in image generation \cite{denton2015deep}, image-to-image translation \cite{isola2017image} and style transfer \cite{johnson2016perceptual}. More recently, unpaired image-to-image translation was successfully learned by adopting a variant of GAN, called cycle-consistent adversarial networks \cite{zhu2017unpaired,taigman2016unsupervised}.
We adopt the concept of CycleGAN for performing the task of non-parallel speech-to-speech 
emotion conversion.


We use a CycleGAN to model the transformation of perturbed speech features $(x \in X)$ to normal speech features $(y \in Y)$.
The CycleGAN model architecture, considered in this work, is motivated from \cite{kaneko2017parallel}.
A typical GAN tries to minimize the adversarial loss $\mathfrak{L}_{adv}(G_{{X \rightarrow Y}}(x), y)$ which measures how far is the generated data $G_{X \rightarrow Y}(x)$ from the target data $y$.
In case of perturbed speech to normal speech transformation without parallel utterances, a typical GAN with only the adversarial loss may not be able to preserve the context information in the speech features.
The CycleGAN model can handle this using a pair of GANs with two adversarial loss functions and an additional cycle consistency loss function.

The first adversarial loss, given as:
\begin{equation}
\mathfrak{L}_{adv}(G_{{X \rightarrow Y}}(x), y)
\end{equation}
corresponds to the forward mapping, which is the transformation from the perturbed speech to normal speech. The second adversarial loss, given as:
\begin{equation}
\mathfrak{L}_{adv}(G_{{Y \rightarrow X}}(y), x)
\end{equation}
corresponds to the inverse mapping, which transforms the normal speech back to the perturbed speech. 

The cycle consistency loss given as:
\begin{align}
\mathfrak{L}_{cyc} = &E_x\left \|G_{Y \rightarrow X}(G_{X \rightarrow Y}(x)) - x\right \|_1 \notag \\
+\ &E_y\left \|G_{X \rightarrow Y}(G_{Y \rightarrow X}(y)) - y\right \|_1
\end{align}
helps to preserve the context information, by ensuring that normal speech can be reconstructed by the cascade of the forward and inverse mapping generators and perturbed speech can be reconstructed by the cascade of the inverse and forward mapping generators, respectively.

In addition to the above mentioned losses, we also included the identity-loss function \cite{zhu2017unpaired}, given as:
\begin{equation}
\mathfrak{L}_{id} = E_x\left \|G_{Y \rightarrow X}(x) - x\right \|_1 
+\ E_y\left \|G_{X \rightarrow Y}(y) - y\right \|_1
\end{equation}
$\mathfrak{L}_{id}$ was originally used for color preservation and  we found this loss to be crucial for maintaining the linguistic information during conversion of speech. 

The complete loss function ($\mathfrak{L}$) of our CycleGAN model is given as:
\begin{align}
\mathfrak{L} = &\mathfrak{L}_{adv}(G_{{X \rightarrow Y}}(x), y) + \mathfrak{L}_{adv}(G_{{Y \rightarrow X}}(y), x) \notag \\ + &\lambda_{cyc} \; \mathfrak{L}_{cyc} + \lambda_{id} \; \mathfrak{L}_{id}
\end{align}
The cycle consistency loss $\mathfrak{L}_{cyc}$ is scaled with a trade-of parameter $\lambda_{cyc}$ whereas the identity-loss $\mathfrak{L}_{id}$ is scaled with a trade-of parameter $\lambda_{id}$.

The generator and discriminator networks in our CycleGAN model consist of different convolutional blocks as shown in Figure \ref{fig:CGAN_Architecture}(a). Gated convolutional (Gated C) blocks consist of gated linear units, which achieved state-of-the-art performance in language and speech modeling, as an activation function for the convolutional layers \cite{dauphin2017language}. Instance normalized gated convolution (I-Gated-C) block uses instance normalization, proposed for style-transfer in \cite{johnson2016perceptual}, after gated C block. Residual convolution (Res-C) blocks are considered to stack multiple convolutional layers, enabling to build a very deep network for the generator \cite{he2016deep}.

The generator network consists a total of $12$ convolutional blocks as shown in Figure \ref{fig:CGAN_Architecture}(b). These include one stride-$1$ gated convolution block, two stride-$2$ I-gated convolution blocks, $6$ residual blocks \cite{he2016deep}, two $\frac{1}{2}$-stride SI-gated convolution blocks, and one stride-$1$ convolution block. All convolution layers are $1$-dimensional to preserve the temporal structure \cite{kaneko2017sequence}. 
The discriminator network consists of $4$ $2$-dimensional convolutional blocks as shown in Figure \ref{fig:CGAN_Architecture}(c). Gated linear units were used as the activation function for all the convolutional blocks. For the discriminator network, we use a $6\times 6$ patch GAN \cite{ledig2017photo, li2016precomputed}, which classifies whether each $6\times 6$ patch is real or fake (i.e., perturbed or normal speech).

\section{Experiments and Results}
\label{exp}
\subsection{Dataset}
We use two spontaneous speech datasets, namely, AMI meeting corpus \cite{mccowan2005ami} and Buckeye corpus of conversational speech \cite{pitt2007buckeye} to analyze the effect of natural perturbations. Both these datasets consist of manual annotations and time-stamps for speech perturbed with emotions and voice-quality. From both these datasets, speech data comprising of $40$ female speakers and $30$ male speakers was considered for training gender-dependent CycleGAN models. We consider $210$ utterances for each gender and for each class (i.e., normal speech, laughter-speech and creaky-speech). Out of these $210$ utterances, $150$ utterances are used for train and $60$ utterances for test. It is to be noted that all these utterances are non-parallel. Each utterance is of $1$-$2$ second(s) in duration.

\begin{table*}
	\centering
	\caption{ASR performance without front-end (no FE) and with front-end (FE). Numbers in parenthesis with $\downarrow$ denote reduction in the error rate.}
	\label{perf_vals}
	\vspace{0.2cm}
	\begin{tabular}{|c|c|c|cc|c|cc|c|cc|}
		\hline
		&&\multicolumn{3}{c|}{Google}&\multicolumn{3}{c|}{IBM}&\multicolumn{3}{c|}{ASpIRE}\\
		\cline{3-5} \cline{6-8} \cline{9-11}
		& & \multirow{2}{*}{no FE} & \multicolumn{2}{c|}{FE} & \multirow{2}{*}{no FE} & \multicolumn{2}{c|}{FE} &\multirow{2}{*}{no FE} & \multicolumn{2}{c|}{FE} \\
		\cline{4-5} \cline{7-8} \cline{10-11}
		& & & MFBs & MFBs+APs& & MFBs & MFBs+APs & & MFBs & MFBs+APs \\ \hline \hline
		Laughter-& \%WER &38.4 &30.9&\textbf{23.5}\;(14.9$\downarrow$) & 50.4 &49.6& \textbf{42.4}\;(8.0$\downarrow$)& 53.5 &45.1& \textbf{32.5}\;(21.0$\downarrow$) \\
		\cline{2-11}
		Speech& \%SER & 91.8 &79.6& \textbf{75.5}\;(16.3$\downarrow$) & 93.1 &89.7& \textbf{89.7}\;(3.4$\downarrow$) &93.1 &91.4& \textbf{89.7}\;(3.4$\downarrow$)\; \\ \hline \hline
		Creaky-& \%WER & 27.4 &22.9& \textbf{16.4}\;(11.0$\downarrow$) &  29.2 &24.3& \textbf{21.3}\;(7.9$\downarrow$) &  32.2 &30.2& \textbf{24.3}\;(7.9$\downarrow$)\; \\
		\cline{2-11}
		Speech& \%SER & 86.1 &77.8& \textbf{63.9}\;(22.2$\downarrow$) & 88.9 &86.1& \textbf{86.1}\;(2.8$\downarrow$)& 94.4 &91.7& \textbf{83.3}\;(11.1$\downarrow$) \\
		\hline
	\end{tabular}
	\vspace{0.2cm}
\end{table*}

\subsection{Feature Extraction}
The WORLD vocoder system \cite{morise2016world} is used to extract features from the speech signal. The speech signal is sampled at $16$ kHz, and  Mel filterbank (MFB) features, logarithmic fundamental frequency ($log$ $F_0$) and aperiodic components (APs) are extracted within a window of length $20$ msec for every $5$ msec. $24$-dimensional MFBs and $24$-dimensional APs are modeled by the proposed CycleGAN architecture to convert the features extracted from the input perturbed speech into the features corresponding to normal speech. Previous work on speaker conversion \cite{ohtani2006maximum, kaneko2017parallel}, have used only the spectral features (MFBs). But for perturbed speech conversion, we found that modeling both, spectral features (MFBs) and aperiodic components (APs) resulted in better conversion to normal speech than considering only spectral features (MFBs). Logarithm Gaussian normalized transformation \cite{liu2007high} was used to convert the $F_0$ values from the source speech to those corresponding to the target speech.

\begin{table}
	\centering
	\caption{DeepSpeech model performance without front-end (no FE) and with front-end (FE) in terms of character error rate (\%CER). Numbers in parenthesis with $\downarrow$ denote reduction in the error rate.}
	\label{perf_DS}
	\vspace{0.15cm}
	\begin{tabular}{|c|c|cc|}
		\hline
		& \multirow{2}{*}{no FE} & \multicolumn{2}{c|}{FE} \\
		\cline{3-4}
		Perturbation& & MFBs & MFBs+APs \\
		\hline \hline
		Laughter speech&56.5 &53.0& \textbf{41.7}\;(14.8$\downarrow$) \\ \hline \hline
		Creaky speech& 33.5 &29.8& \textbf{23.7}\;(9.8$\downarrow$) \\ \hline
	\end{tabular}
	\vspace{0.1cm}
\end{table}

\subsection{Training Details}
In order to achieve a more stable training of the CycleGAN models and to generate higher quality outputs, we used the least square function to compute the adversarial loss instead of the commonly used negative log likelihood objective function \cite{mao2017least, zhu2017unpaired}. The CycleGAN models were trained using the Adam optimizer with a batch size of $1$. The initial learning rates of the generator and the discriminator are $0.0002$ and $0.0001$, respectively. The learning rates were decayed by a factor of $10^5$ after each epoch. In all the experiments, the cycle consistency loss trade-of parameter $\lambda_{cyc}$ was set to a value of $10$. The identity-loss trade-of parameter $\lambda_{id}$ was set to $1$ for the first $100$ epochs and set to $0$ after $100$ epochs.

In this paper, we have trained gender-specific CGAN models (using speech collected from multiple speakers within the same gender) to transform perturbed speech to normal speech. But unlike the task of voice conversion, where each model is trained to convert speech between a pair of speakers, our proposed CGAN models are able to handle variations across multiple speakers (within the same gender) to transform perturbed speech to normal speech. Moreover, we have considered speech from speakers unseen during training to test the scalability of the trained CGAN models.

\begin{table}
	\centering
	\caption{Outputs of ASpIRE ASR model when different signals (original and  transformed) are provided as input. Note: Gnd. and orig. refers to ground truth and original signal, respectively. MFBs, (MFBs+APs) refers to signals transformed using CGAN models trained on only MFBs and MFBs + APs, respectively.}
	\label{ASpIRE_outputs}
	\begin{tabular}{|l|l|}
		\hline
		\multicolumn{2}{|c|}{(a) Laughter-speech} \\ \hline
		Gnd.&see if anything out there to see you such \\ \hline
		\multirow{2}{*}{Orig.}&[noise] see anything out there to you \\
		&[laughter] \\ \hline
		MFBs &see if anything out there to you use \\ \hline
		MFBs+APs &see if anything out there to see you \\ \hline \hline
		\multicolumn{2}{|c|}{(b) Creaky-speech} \\ \hline
		Gnd.& think they should get rid of it\\ \hline
		orig.&like they should get kid \\ \hline
		MFBs &think they should get rid of \\ \hline
		MFBs+APs &think they should get rid of \\ \hline
	\end{tabular}
\end{table}

\begin{figure*}[htb]
	\vspace{0.3cm}
	\centering
	\begin{subfigure}[b]{0.33\textwidth}
		\centering
		\includegraphics[width=\linewidth]{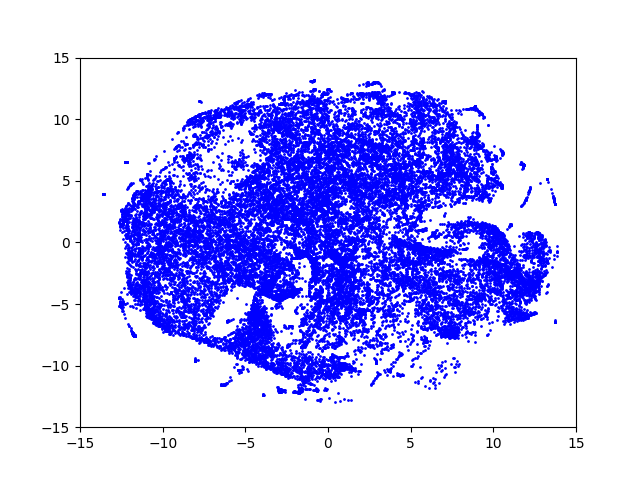}
		\caption{Normal Speech}
		\label{fig:tsne-ns}
	\end{subfigure}\hfill
	\begin{subfigure}[b]{0.33\textwidth}
		\centering
		\includegraphics[width=\linewidth]{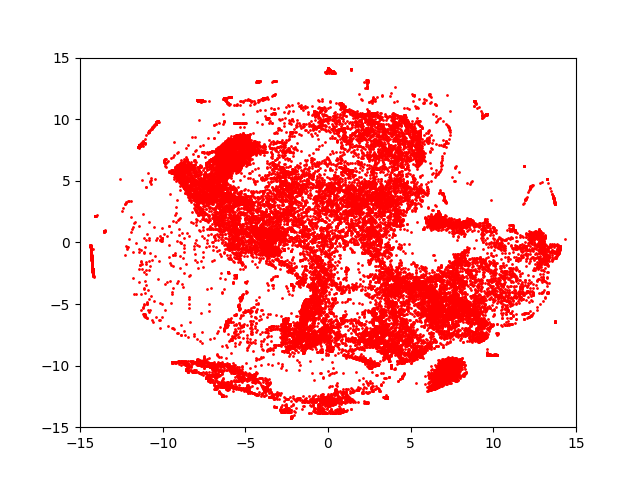}
		\caption{Laughter Perturbed Speech}
		\label{fig:tsne-ls}
	\end{subfigure}\hfill
	\begin{subfigure}[b]{0.33\textwidth}
		\centering
		\includegraphics[width=\linewidth]{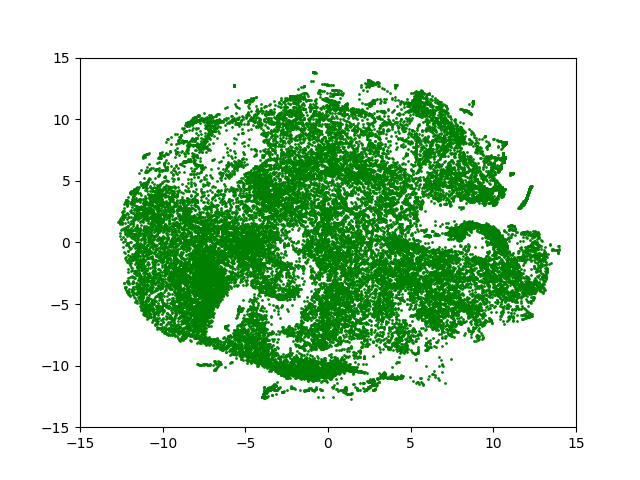}
		\caption{Transformed (Normal) Speech}
		\label{fig:tsne-nsc}
	\end{subfigure}
	\caption{t-SNE projection of Mel filterbank output features (Best viewed in color).}\label{fig:tsne}
\end{figure*}

\subsection{Results}
Table \ref{perf_vals} presents the performance of Google cloud ASR\footnote{https://cloud.google.com/speech-to-text/}, IBM ASR\footnote{https://www.ibm.com/watson/services/speech-to-text/} and Kaldi ASR (with ASpIRE models) \cite{peddinti2015jhu, aspirelink} with and without our proposed front-end, when tested with laughter-speech (speech perturbed with emotion) and creaky-speech (speech perturbed with voice-quality). The performance is evaluated in terms of \% Word Error Rate (\%WER) and \% Sentence Error Rate (\%SER). Lower values of WER and SER indicate better performances.
Table \ref{perf_vals} shows that our proposed front-end improves the performance of each of the ASR systems.
It can be observed that modeling both, spectral and aperiodic components (i.e., MFB + APs) performs better than modeling only MFBs in the proposed front-end. Absolute reductions of $14.9$\%, $8$\%, $21.0$\% in WER and $16.3$\% , $3.4$\%, $3.4$\% in SER, are achieved for Google ASR, IBM ASR and ASpIRE ASR, respectively, when our proposed front-end (MFBs+APs) is used to convert laughter-speech to normal speech. Similarly, absolute reductions of $11.0$\%, $7.9$\%, $7.9$\% in WER and $22.2$\%, $2.8$\% $11.1$\% in SER are obtained for Google ASR, IBM ASR and ASpIRE ASR, respectively, when our proposed front-end (MFBs+APs) is used to convert creaky-speech to normal speech.

The ASR performances shown in Table \ref{perf_vals} are influenced by the strength of the language model used by the respective ASR systems. To check ASR performance without the effect of a language model, we also present the results from the DeepSpeech model\footnote{https://github.com/mozilla/DeepSpeech/releases/tag/v0.4.0-alpha.3} which converts speech to a sequence of English characters. Table \ref{fig:box} shows the \% Character Error Rate (\%CER) performance of the DeepSpeech model with and without the proposed front-end. The DeepSpeech model was trained on $1000$ hours of LibriSpeech data and did not use a language model for decoding. It can be observed from Table \ref{fig:box} that our proposed front-end gives significant reduction in CER of the DeepSpeech model.

Table \ref{ASpIRE_outputs} shows the speech-to-text outputs of an ASpIRE model when tested with original (orig.) perturbed speech (laughter-speech and creaky-speech), transformed normal speech samples using CGAN models trained using MFBs and MFBs+APs, respectively. For instance, it can be observed from Table \ref{ASpIRE_outputs} that the original laughter-speech signal is transcribed as having [noise] and [laughter] along with substitutions and deletions (i.e., if, see and such) whereas the CGAN transformed signals have no [noise] and [laughter] (for both MFBs and MFBs+APs)
and has less deletions (to, use and such for MFBs, and only such deleted for MFBs+APs). Similarly for creaky-speech as input, CGAN transformed signals are better transcribed (only 'it' is deleted for both MFBs and MFBs+APs models) compared to the original (orig.) creaky-speech signal ("of it" is deleted and "rid" is substituted with "kid" for the orig. signal).

\section{Analysis of the Learned Front-end Transformation}
\label{analysis}
Figure \ref{fig:tsne} shows a $2$-dimensional t-SNE projection \cite{vanDerMaaten2008} of the Mel filterbank features for (a) normal speech, (b) laughter perturbed speech \cite{dumpala2014analysis} and (c) laughter perturbed speech transformed to normal speech by the proposed front-end.
It can be observed that the filterbank features for normal speech and transformed (normal) speech are quite similar to each other and that they differ significantly from the filterbank features for laughter-speech. Additionally, the spread of the filterbank features for laughter-speech is reduced in the $2$-dimensional t-SNE space. We hypothesize that this may be due to the reduction in vowel space for laughter-speech \cite{bachorowski2001acoustic}.

\begin{figure}
 	\centering
 	\includegraphics[width=\linewidth]{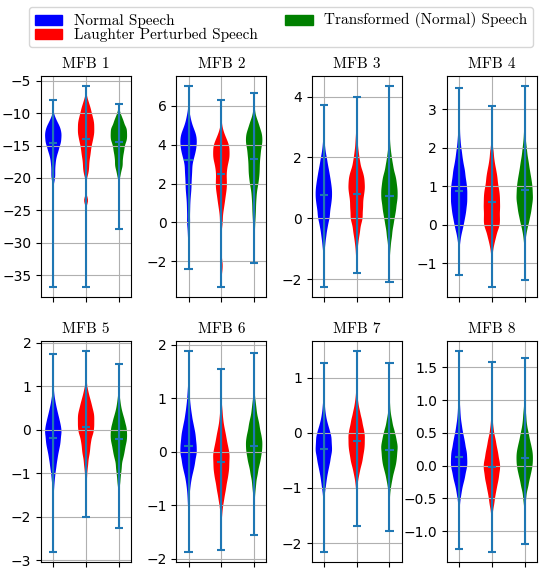}
 	\caption{Violin plot of output from filters $1$ to $8$  of the Mel filterbank (Best viewed in color).}\label{fig:box}
\end{figure}

For a more detailed analysis, Figure \ref{fig:box} shows violin plots \cite{Hintze.Nelson1998} of the output of the filters $1$ to $8$ of the Mel filterbank, for normal speech, laughter perturbed speech and laughter perturbed speech transformed to normal speech. Output of the filters $9$ to $24$ do not show visible differences and hence they are not shown. It can be observed from Figure \ref{fig:box} that the distribution of the feature values for normal speech and transformed (normal) speech are similar and they exhibit similar variations. 
It implies that the front-end is able to (a) capture the distribution of the Mel filterbank outputs of both normal and laughter perturbed speech, and (b) transform  laughter perturbed speech to equivalent normal speech.

\section{Conclusion}
\label{conc}
We proposed a novel front-end based on CycleGANs to transform naturally perturbed speech to normal speech. Experiments on spontaneous laughter-speech and creaky-speech utterances show significant improvements in performance of the Google ASR, IBM ASR, the Kaldi ASR with ASpIRE model and that of a DeepSpeech model. We found that adding aperiodic components to spectral features gives a better performance. Visualization of the laughter-speech features and the transformed  speech features gives insights on the transformation performed by our proposed front-end.

\bibliographystyle{IEEEbib}

\bibliography{references_mod.bib}

\end{document}